# SPECIFIC HEAT OF THE 38-K SUPERCONDUCTOR MGB$_2$ IN THE NORMAL AND SUPERCONDUCTING STATE: BULK EVIDENCE FOR A DOUBLE GAP.

Alain JUNOD, Yuxing WANG, Frédéric BOUQUET, Pierre TOULEMONDE

*Département de Physique de la Matière Condensée, Université de Genève,*
*CH-1211 Genève 4 (Switzerland)*

## 1. Introduction.

Specific heat has traditionally been regarded as a critical test experiment in the development of the theory of superconductivity, and in the quest for new superconducting materials. The magnitude of the specific heat jump at the transition temperature $T_c$, and the exponentially vanishing specific heat at low temperature, unveiling the existence of a gap in the spectrum of electronic excitations, contributed to establish the validity of the BCS theory. The detection of an anomaly at $T_c$ in new or unconventional superconductors is considered as one of few proofs of bulk superconductivity. Therefore, shortly after the discovery of superconductivity in MgB$_2$ near 40 K [1], several groups rapidly performed specific heat measurements [2] [3] [4] [5] [6] [7]. The present discussion is based essentially on Geneva data [5] extending over a wide field and temperature range, but we underline a good agreement with results of groups using high magnetic fields to determine the normal-state behavior [5] [6] [7]. Up to now only polycrystalline samples have been investigated. Noticeably, the Berkeley experiments were performed on an isotopically pure specimen [6] and are discussed elsewhere in this volume [8].

The discovery of high temperature superconductivity (we may call "high $T_c$" any $T_c$ substantially higher than the long-lasting record of 23 K in the A15-phase Nb$_3$Ge) in a simple binary compound has again put into question the limits of $T_c$ for superconductors having an electron-phonon interaction providing the coupling to form Cooper pairs.

Traditional evaluations set a limit of ≈ 30 K for this class of superconductor [9]. The origin of the superconducting coupling in cuprates, with critical temperatures as high as 160 K, is still controversial, and may be of magnetic origin. In contrast, the observation of a large isotope effect in $MgB_2$ has provided strong evidence in favor of a phonon-mediated coupling [2]. The question of the origin of high $T_c$ in $MgB_2$ has been then tackled by modern computational methods based on extensive band-structure calculations [10] [11] [12] together with selective coupling to specific phonon modes [13] [14] [15]. In this work, using an experimental approach, we compare average phonon and electron quantities for $MgB_2$ with other representative superconductors such as $Nb_3Sn$ and $YBa_2Cu_3O_7$, to pin-point the essential anomalies of $MgB_2$ in the framework of phenomenological $T_c$ equations. Anticipating the results, these uncommon characteristics are: a small electronic density of states at the Fermi level, high phonon frequencies, and a two-gap structure.

The paper is organized as follows: Section 2 reviews the links between the superconducting transition temperature and average material parameters, in particular those which are accessible to specific heat experiments, and lists the questions that will be addressed. In Section 3, samples are described, and their magnetic properties are characterized in Section 4. Specific heat measurements are presented in Section 5-7 in the high, medium, and low temperature range, respectively. The phenomenological two-gap model introduced in Section 6 is a central point of this paper. The anomalous field dependence at low temperature is detailed in Section 8. Ginzburg-Landau parameters are evaluated in Section 9. Technical details are given in the appendices.

## 2. BCS background.

The BCS theory and its subsequent refinements based on the Eliahsberg equations show that high critical temperatures in phonon-mediated superconductors are favored by high phonon frequencies, and by a large density of states at the Fermi level. The Allen-Dynes ("AD") formula [16], an interpolation based on numerical solutions of the Eliashberg equations valid over a wide range of coupling strengths and for various shapes of the phonon spectrum, expresses quantitatively:

$$T_c = \frac{\langle \omega_{\ln} \rangle}{1.20} \exp\left(-\frac{1.04(1+\lambda_{ep})}{\lambda_{ep} - \mu^* - 0.62\lambda_{ep}\mu^*}\right)$$

$$\lambda_{ep} \equiv 2\int_0^\infty \alpha^2 F(\omega) \, d\ln\omega = \frac{N(0)\langle I^2 \rangle}{M\overline{\omega}_2^2}$$

$$\overline{\omega}_2 \equiv \left(\frac{2}{\lambda_{ep}} \int_0^\infty \omega^2 \, \alpha^2 F(\omega) \, d\ln\omega \right)^{1/2}$$

$$\omega_{\ln} \equiv \exp\left(\frac{2}{\lambda_{ep}} \int_0^\infty \ln\omega \, \alpha^2 F(\omega) \, d\ln\omega \right)$$

where $\lambda_{ep}$ is the dimensionless electron-phonon coupling, $\mu^* \cong 0.1$ is the retarded Coulomb repulsion, $F(\omega)$ is the normalized phonon density of states (PDOS), $\alpha^2 F(\omega)$ is the spectral electron-phonon interaction function, $N(0)$ is the density of states at the Fermi level per spin and per atom (EDOS), $<I^2>$ is the properly averaged electron-ion matrix element squared, and $M$ is the average atomic mass. $N(0)<I^2> \equiv \eta$ is the so-called Hopfield electronic parameter, whereas $M\overline{\omega}_2^2$ is an average force constant. Correction factors for very strong coupling ($\lambda_{ep} > 3$) are omitted. The essential difference with McMillan's equation [9] is the use of $\omega_{ln}$ rather than the Debye temperature as a characteristic phonon energy.

Specific heat ($C$) measurements give access to several of these parameters. The specific heat jump at $T_c$ provides a measurement of the bulk critical temperature. Unlike resistivity measurements, sensitive to the first percolation path, and susceptibility measurements, sensitive to shielding by a superficial layer of superconducting material, the specific heat jump is proportional to the superconducting volume, and its width mirrors the distribution of $T_c$ within the whole volume.

The Sommerfeld constant $\gamma = \lim_{T \to 0} C_n / T$ is another important parameter given by the specific heat. $C_n(T)$, the normal-state specific heat, can be measured using a field larger than the upper critical field $H_{c2}(0)$ (if available). Furthermore, $\gamma$ is proportional to the EDOS renormalized by the electron-phonon interaction:

$$\gamma = \frac{2}{3}\pi^2 k_B^2 N(0)(1+\lambda_{ep})$$

where $k_B$ is the Boltzmann constant and $N(0)$ is the EDOS for *one* spin direction. Expressing $N(0)$ in states per eV per atom and per spin direction, and $\gamma$ in mJ/K$^2$gat (gat $\equiv$ gram-atom), this formula becomes $N(0)(1+\lambda_{ep}) = 0.212\gamma$.

The lattice specific heat provides information on the phonons. The slope of the specific heat, $\beta_3 = \lim_{T \to 0} d(C/T)/d(T^2)$, gives the initial Debye temperature

$$\Theta_0 = \left(\frac{12}{5}\frac{R\pi^4}{\beta_3}\right)^{1/3}$$

which depends on the sound velocities; $R$ is the ideal gas constant. In practical units, $\Theta_0 = (1944000/\beta_3)^{1/3}$ if $\beta_3$ is given in mJ/K$^4$gat and $\Theta_0$ in K. Deviations from the Debye model are common. MgB$_2$ is no exception, the initial Debye temperature is generally not representative of the frequencies that are most effective for pairing (i.e., typically $\omega = 2\pi T_c$ [17]).

More representative averages can be defined if $F(\omega)$ is known. The spectral coupling function $\alpha^2 F(\omega)$, entering in AD's formula, is not directly available without e.g.

tunneling experiments. As a first approximation, $F(\omega)$ is sufficient assuming uniform coupling, i.e. $\alpha^2 F(\omega) \propto F(\omega)$:

$$\overline{\omega}_2 \cong \left(\langle\omega\rangle/\langle\omega^{-1}\rangle\right)^{1/2}$$
$$\omega_{\ln} \cong \exp\left(\langle\omega^{-1}\ln\omega\rangle/\langle\omega^{-1}\rangle\right)$$
$$\langle\phi(\omega)\rangle \equiv \int_0^\infty \phi(\omega)F(\omega)d\omega$$

In this approximation, $\langle\alpha^2\rangle = 0.5\lambda_{ep}/\langle\omega^{-1}\rangle$. The moments $\langle\omega^n\rangle$ can be obtained by an inversion procedure of the specific heat curve [18] [19], or equivalently from particular integrals of the lattice specific heat $C_{ph}(T)$ [20] [21]. This follows from the fundamental relation between the phonon specific heat and the PDOS:

$$C_{ph}(T) = 3R\int_0^\infty E(\omega/T)F(\omega)d\omega,$$
$$E(x) \equiv x^2 e^x (e^x+1)^{-2}$$

where $E(x)$ is the Einstein specific heat in units of $k_B$ for one mode with frequency $\omega$ at the temperature $T$, and $x \equiv \omega/T$ (we express frequencies in Kelvin, omitting the implicit factor $\hbar/k_B$). Although finding the precise PDOS from specific heat data is an ill-conditioned problem, the determination of a smoothed PDOS or moments of the PDOS is well-conditioned. As a rule of thumb, in order to sample the full phonon spectrum up to a frequency $\omega$, one needs specific heat data for temperatures up to $T \approx 0.2\omega$. The lattice specific heat must be isolated from other contributions. For superconductors, the preferred method consists of measuring the normal-state specific heat at high magnetic fields, and then to subtract the electronic contribution determined at $T \to 0$. Corrections due to the difference between the constant-volume and constant-pressure specific heat, the temperature-dependence of $\gamma$ or its renormalization are generally small and can be neglected.

Specific heat measurements then give information on the electron-phonon coupling strength. The ratio $T_c/\omega_{\ln}$ and the AD formula determine $\lambda_{ep}$. The dimensionless specific heat jump $\Delta C/\gamma T_c$ also depends on $\lambda_{ep}$, and varies from 1.43 in the weak-coupling limit (e.g. for aluminium with $T_c$ = 1.2 K) to about 3 for strong-coupling superconductors (e.g. La$_3$In with $T_c$ = 9.5 K [22]). The phenomenological model of Padamsee *et al.* [23] gives a quantitative description for the influence of the coupling strength on the thermodynamics through a single parameter, $2\Delta(0)/k_B T_c$; $\Delta(0)$ is the superconducting gap at $T = 0$. The ratio $2\Delta(0)/k_B T_c$ varies from the BCS value 3.53 to over 5 for La$_3$In [22] and YBa$_2$Cu$_3$O$_7$ [24]. Table II recalls the characteristic parameters for Nb$_3$Sn and YBa$_2$Cu$_3$O$_7$, and also lists the essential results obtained in this work for MgB$_2$.

One should keep in mind that McMillan's or AD's equations are interpolation formulas, which represent the solutions of *isotropic* Eliashberg equations. An anisotropic gap $\Delta(\vec{k})$ distributed between $\Delta_{min}$ and $\Delta_{max}$ could result in a smaller specific heat jump, and simulate weaker coupling.

We performed specific heat experiments in order to address the following questions:

- how large are the *average* phonon frequencies? Are they large enough to account for $T_c$ via the phononic prefactor $\omega_{ln}$?
- how large is the EDOS $N(0)$? Is it large enough to balance the phononic denominator $M\overline{\omega}_2^2$, and to produce a large value of $\lambda_{ep}$?
- is MgB$_2$ a strong- or weak-coupling superconductor?
- how does the electron-phonon interaction compare with other superconductors?
- is the condensation energy $E_c$ unusually large?
- is there any indications in favor of a $\vec{k}$ and band dependence of the gap?
- what are the values of the Ginzburg-Landau parameters $\lambda$, $\xi$, $\kappa$?

For this purpose, we measured the specific heat in the normal state down to 3 K in $B$ = 14 and 16 T. The latter curves are undistinguishable, so we believe that these fields are very close to or larger than $H_{c2}$. This allowed us to determine the Sommerfeld constant with a small extrapolation error. We measured the magnetic susceptibility in order to provide an independent estimation of the EDOS. In zero field, we measured the specific heat from 2 to 300 K. Using a simple inversion technique, we determined the moments of the PDOS. We applied intermediate fields in various temperature ranges, to measure the shift of the critical temperature and the variation of the low temperature specific heat. This provided a bulk determination of the initial slope of $H_{c2}$, and unveiled an unusual behavior of $C$ at $T \ll T_c$ that differed strongly from an isotropic s-wave type-II superconductor both in the $T$- and $B$- dependence. We conclude that $T_c$ is enhanced both by high phonon frequencies and by a complex gap structure in MgB$_2$.

## 3. Samples

Samples synthesized by two different routes were investigated. For the first series, commercial powder of MgB$_2$ (Alfa Aesar), with a nominal 98% purity, was pressed into pellets, sealed into quartz tubes under 1 bar of argon gas at ambient temperature, and sintered for 60 hours at 850°C. No measurable weight losses occurred. One cylindrical pellet with a diameter/height ratio of ≈ 1 and mass of 0.1 gram was used for the magnetic measurements, and one disc-shaped pellet with a mass of 0.15 gram was used for the specific heat measurements (sample code "A2"). X-ray diffraction showed only lines of the AlB$_2$ structure, together with one faint unidentified line. Iron impurities were detected by microprobe analysis at the 1-2% level, in addition to some carbon and oxygen.

The second batch was synthesized from the different elements. Powders of Mg (99.8%) and B (99.7%) (Alfa Aesar) were mixed, compacted into a boron nitride crucible, and sealed in a furnace inside a 500 tons cubic press. The sample was reacted for one hour at 900°C and 3 Gpa (sample code "HP14"). A pellet with a mass of 0.3 gram was used for the high temperature specific heat measurements (15-120 K), then cut to 0.1 and 0.05 gram samples used for the low temperature specific heat (1.5-15 K) and magnetic measurements respectively. X-ray diffractograms revealed the $AlB_2$ structure, and an estimated 5% of MgO. SEM and microprobe analysis additionally detected a small amount of an amorphous $MgB_{2+x}$ boron-rich phase.

## 4. Magnetic properties

AC susceptibility, at 8 kHz and 0.1 G, shows an onset of superconductivity at 38 K for sample A2 (Fig. 1a). In the 30-10 K range, a broad increase of diamagnetism accompanied with a signal in the quadrature (loss) channel indicate that weak links in-between grains become superconducting, expelling the field from of the entire volume of the porous pellet. Sample HP14 is free of weak links, and shows a single transition centered at 37 K (Fig. 1b). Note that the weak link "transition", commonly seen in e.g. $Bi_2Sr_2CaCu_2O_8$ ceramics, is not a thermodynamic property and never gives rise to any features in specific heat curves.

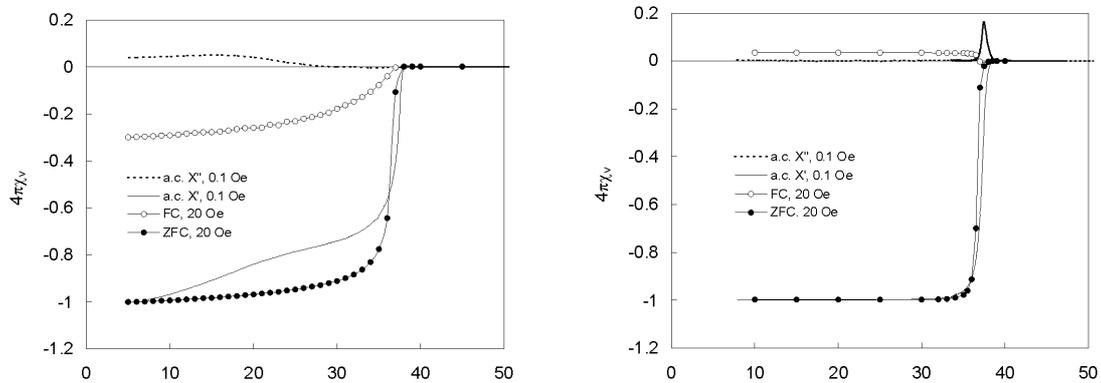

*Figure 1. (a)*, left panel: diamagnetic transition of sample A2. Full line: AC susceptibility, in-phase $\chi'$ component. The signal is normalized to $-1$ at low temperature. Dotted line: quadrature (loss) component $\chi''$. (●), zero field cooled (ZFC) susceptibility. (○), field cooled (FC) Meissner susceptibility. The DC susceptibility in $emu/cm^3$ is corrected for a demagnetization factor $N = 1/3$. *(b)*, right panel: same plot for sample HP14.

The DC susceptibility of sample A2, measured in a SQUID magnetometer in a field of 20 Oe (zero field cooling conditions), shows full diamagnetism at low temperature ($\chi = -1/4\pi$), taking into account the demagnetization factor $N \cong 1/3$ for the short cylinder (Fig. 1a). The normalization volume for this experiment is the measured mass divided by the density as evaluated using measured lattice parameters, 2.625 g/cm$^3$. 60% of the transition occurs over a 1 K interval, with an onset between 37 and 38 K. The Meissner curve (field cooling conditions, 20 Oe) shows an onset at 37 K, and 30% diamagnetism at low temperature.

The DC susceptibility of sample HP14 shows two prominent differences (Fig. 1b). The field cooling Meissner magnetization is not only much smaller in absolute values, indicating stronger flux pinning, but also *paramagnetic*. This anomaly was also observed in high density samples made of Alfa MgB$_2$ powder after hot-pressing at 950-1150 °C. The latter samples have almost ideal macroscopic density, and exhibit sharp transitions with $\Delta T_c \approx 1$ K [25]. However their calorimetric transition (not shown) is slightly broader than for sample A2. The occurrence of a paramagnetic Meissner effect in a *s*-wave superconductor ("Wohlleben effect") was studied in detail in Nb, and attributed to inhomogeneously trapped flux [26].

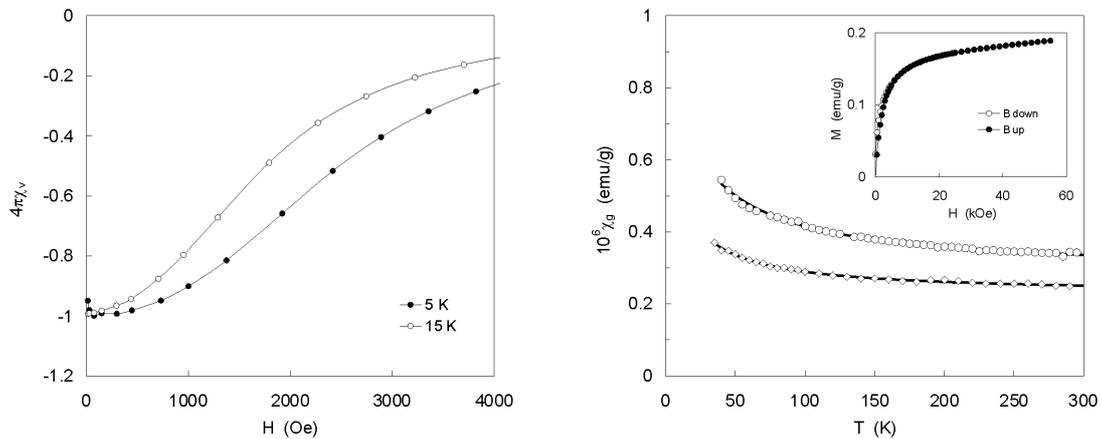

*Figure 2. (a)*, left panel: susceptibility $4\pi M / H$ of sample A2 in the superconducting state versus field at 5 and 15 K, virgin curves. *(b)*, right panel: normal-state susceptibility of sample A2 (○) and HP14 (◊) versus temperature. The lines through the data are fits with Pauli and Curie-Weiss terms. *Inset*: magnetization cycle of sample A2 at T = 50 K showing the saturation of Fe impurity moments.

Magnetization cycles in the superconducting state indicate strong pinning, even for sample A2 that was sintered at a relatively low temperature [5]. The first departure from full diamagnetism occurs between 300 and 500 Oe at 5 K, and sets an upper limit to $H_{c1}(0)$ (Fig. 2a).

Figure 2b shows the normal-state susceptibility between 40 and 300 K for sample A2. The raw magnetization data contain a ferromagnetic contribution, as shown in the inset, which saturates above ≈ 2 teslas. Its amplitude is equivalent to the contribution of 730 ppm *ferromagnetic* Fe clusters. The susceptibility is obtained from the high field slope:

$$\chi = \frac{M(H_2) - M(H_1)}{H_2 - H_1}$$

with $\mu_0 H_2 = 5.5$ T and $\mu_0 H_1 = 3$ T. Its variation is well described by $\chi = \chi_0 + C_c/(T - \Theta_c)$, where $\chi_0 = 4.4\times10^{-6}$ emu/gat, $\Theta_c = -26$ K, and $C_c = 2.5\times10^{-4}$ emu·K/gat (equivalent to 58 ppm *paramagnetic* diluted Fe ions). By subtracting $\chi_c = -1\times10^{-6}$ emu/gat for the diamagnetic core susceptibility of $Mg^{2+}$ ions [27], one finds $\chi_0 = 5.4\times10^{-6}$ emu/gat for the Pauli susceptibility. This corresponds to an EDOS $N(0) = 0.083$ states/eV·atom·spin, about 30% smaller than what is found from band structure calculations [10] [11] [12] [13]. The unfavorable ratio between the small Pauli susceptibility and the Curie and ferromagnetic contributions (let alone van Vleck and Landau-Peierls terms) leads to a large uncertainty.

For sample HP14 (Fig. 2b), the Fe equivalent of ferromagnetic clusters is 126 ppm; $\chi_0 = 3.5\times10^{-6}$ emu/gat, $\Theta_c = -15$ K and $C_c = 1.06\times10^{-4}$ emu·K/gat (equivalent to 24 ppm paramagnetic Fe). The apparent Pauli susceptibility is probably lowered by the fraction of MgO. For hot-pressed samples (not shown), the total differential susceptibility between 3 and 5.5 T is practically zero, or slightly diamagnetic at all temperatures above 70 K.

Paramagnetic moments cause both a Curie term in the magnetic susceptibility and a Schottky anomaly in the low temperature specific heat. Therefore sample HP14, with a 2.4 times smaller $C_c$ than for sample A2, is preferred for the study of the specific heat as a function of the magnetic field at $T \ll T_c$. Whereas sample A2, containing $\geq$ 98% $MgB_2$ phase, is preferred for zero-field studies and general characterization. The quantitative relation between the apparent Curie contribution in the susceptibility and the Schottky anomaly in the specific heat is not exact, as it may be falsified by a temperature-dependent Pauli susceptibility.

## 5. Specific heat at high temperature: phonons

> *"A metal that does not have a high Debye temperature (as do metallic H, B, Be or their alloys) cannot become a high $T_c$ superconductor" [28].*

Between 15 and 300 K, the specific heat was measured in an adiabatic calorimeter provided with Pt thermometry [29]. The sample was continuously heated at 15 mK/s and the heat capacity $c$ was obtained from $c = P/(dT/dt)$. Residual losses, typically $\pm$ 0.1% of the heater power $P$, were measured every $\approx$ 10 to 20 K and corrected for. The heat capacity of the sample holder and other addenda was measured separately and subtracted. This calorimeter typically gives 0.01% scatter, 0.2% reproducibility and 0.5% accuracy in fields $0 \leq B \leq 16$ T, using 0.3 gram Cu samples. With $MgB_2$, which has a high Debye temperature, an uncommon problem arises at the lower end of the temperature range: the heat capacity of the 1 mg silicone grease used as an adhesive is as large as 20% of the heat capacity of the 150 mg sample. This is the main limitation on the absolute accuracy of the measurement.

Figure 3 shows the data in the 15-300 K temperature region. Data in $B = 0$, 10 and 14 T are shown, and merge into a common line above $T_c$. The specific heat reaches only two third of the Dulong-Petit equipartition value at room temperature. This indicates a high effective Debye temperature, $\Theta_{eff} \cong 920$ K at $T = 298$ K (at these temperatures, $\Theta_{eff}$ is determined by the second moment of the PDOS). Low temperature data (Section 7) give an initial Debye temperature $\Theta_0 \approx 800$ K for sample A2 (average, 3-15 K), and $\Theta_0 = 930 \pm 10$ K for sample HP14. The initial Debye temperature probes only the $\omega \to 0$ part of the PDOS. Sample-dependent variations are usually observed in porous sinters, and the higher value is considered as more representative for the sound velocities. The inset of Fig. 3 shows the PDOS on a logarithmic scale, compared to a Debye PDOS which has the same moment $\langle\omega^2\rangle$. The highest phonon frequencies exceed 1000 K, and there is an excess weight in the vicinity of 330 K, which coincides with an acoustic phonon branch according to neutron studies [30].

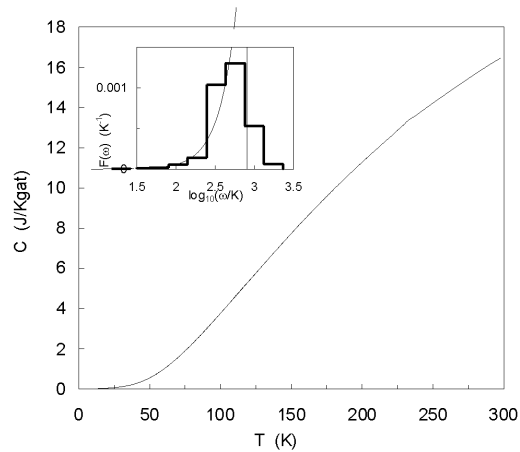

*Figure 3.* *Specific heat of sample A2 from 15 to 300 K. About 8000 data points for B = 0, 10 and 14 T merge into the thickness of the line.* **Inset**: *model of the phonon DOS used to fit the data. The weights $w_i$ are divided by the spacing $0.567\omega_i$ between Einstein frequencies $\omega_i$ to build the histogram. The thin line is a Debye PDOS.*

Figure 4 illustrates how the PDOS is obtained. $F(\omega)$ is represented by nine δ-functions with frequencies in a geometrical ratio 1.75:1, ranging from 20 K to 1760 K. The weights of these Einstein components are independently fitted to the specific heat curve (Table I). The ratio 1.75:1 is chosen to be large enough to avoid spurious oscillations of the PDOS (in other words, to avoid an ill-conditioned problem). The sum of the weights is constrained to 1 ($3N$ modes per gat). Figure 4 shows the contribution of individual frequencies, their sum, and the data points in the normal state in the form $(C - \gamma T)/T^3$, for two values of the Sommerfeld constant that we consider as upper and lower bounds. The upper bound is the lowest measured value of $C/T$, at the lowest measured temperature in $B = 14$ or 16 T, i.e. 0.95 mJ/K$^2$gat at 3 K. The lower bound is an extrapolation of the 14 or 16 T curves to $T = 0$ using a polynomial expansion, i.e. 0.84 mJ/K$^2$gat.

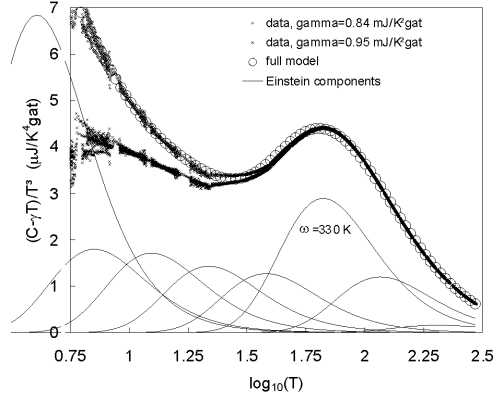

***Figure 4.*** *Plot illustrating the deconvolution of the specific heat into an approximate PDOS. The lattice specific heat of sample A2 is shown in the form $(C_{total} - \gamma T)/T^3$. The lower bound for $\gamma$ is subtracted from the upper data set (♦), the upper bound for $\gamma$ is used for the lower data set (×). The contributions of the individual Einstein frequencies, equally spaced on the logarithmic temperature axis, are shown in the lower part after fitting their weight (Table I) to the upper set of data. Their total contribution is shown by (○).*

One notices the large weight of the acoustic mode near 330 K, consistent with first-principle calculations [15] and neutron experiments [30]. The increasing PDOS at very low energies might be due to a soft phonon mode, however it could be an artifact due to the broad Schottky contribution of isolated Fe ions or to the broad superconducting transition near $H_{c2}$. For an ideal Debye PDOS, $F(\omega) = 3\omega^2/\omega_D^3$ for $\omega < \omega_D$, and $C_{phonon} = \beta_3 T^3$ for $T < \Theta_D/10$, aligning the data on an horizontal line below $\log_{10} T \approx 2$. The deviation from this behavior shows that an extrapolation of the normal-state specific heat, from above $T_c$, into the superconducting state would be unreliable.

***Table I.*** *Position and weight of the Einstein modes used to fit the lattice specific heat of MgB$_2$. The total normal-state specific heat can be reconstructed from the formula C [J/Kgat] = 24.94 $\Sigma_i w_i E(\omega_i/T) + 0.00084T$. Frequencies are in a geometric progression, $\omega_i = (1.75)^i \omega_0$.*

| i | $\hbar\omega_i/k_B$ (K) | $w_i$ |
|---|---|---|
| 0 | 20 | 0.000103 |
| 1 | 35 | 0.000145 |
| 2 | 61.3 | 0.000739 |
| 3 | 107.2 | 0.00331 |
| 4 | 187.6 | 0.0157 |
| 5 | 328.3 | 0.192 |
| 6 | 574.5 | 0.427 |
| 7 | 1005 | 0.300 |
| 8 | 1759 | 0.0602 |

|  | Nb$_3$Sn | YBa$_2$Cu$_3$O$_7$ | MgB$_2$ |
|---|---|---|---|
| $M_{gat}$ [g] | 99.4 | 51.5 | 15.31 |
| $V_{gat}$ [cm$^3$] | 11.1 | 8.0 | 5.83 |
| $T_c$ [K] | 18 | 90 | 36.7 |
| $\Theta_0$ [K] | 230 | 425 | ≈ 800 (A2) <br> 930 ± 10 (HP14) |
| $\omega_{ln}$ [K] | 125 |  | 480 |
| $\overline{\omega}_2$ [K] | 174 |  | 633 |
| $<\omega^2>^{1/2}$ [K] | 226 |  | 808 |
| $\gamma$ [mJ/K$^2$gat] | 13 | ≈ 1.5 | 0.89 ± 0.05 (A2) <br> 0.83 ± 0.02 (HP14) |
| $N(0)$ [eV$^{-1}$atom$^{-1}$spin$^{-1}$] | 0.99 | ≈ 0.13 | 0.12 |
| $\lambda_{ep}$, from $T_c/\omega_{ln}$ | 1.8 |  | 1.07 |
| $\lambda_{ep}$, from $\gamma/N(0)$ | 1.8 | ≈ 1.5 | 0.58 |
| $M\overline{\omega}_2^2$ [eV/Å$^2$] | 5.3 |  | 10.9 |
| $N(0)<I^2>$ [eV/Å$^2$] | 9.5 |  | 6.3 |
| $\Delta C/\gamma T_c$ | 2.5 | 2? | 0.82 |
| $\Delta C/k_B$ [1/cm$^3$] | 3.8×10$^{21}$ | ≈ 2.5×10$^{21}$ | 3.3×10$^{20}$ |
| $\mu_0 H_c(0)$ [T] | 0.52 | ≈ 1.0 | 0.26 |
| $E_c$ [mJ/cm$^3$] | 108 | ≈ 400 | 27 |
| $0.2364\gamma T_c^2$ [mJ/cm$^3$] | 90 | ≈ 360 | 49 |
| $2\Delta(0)/k_B T_c$ | 4.8 | 5 | 1.2-4.2 (*) |
| $-\mu_0(dH_{c2}/dT)_{T_c}$ [T/K] | 1.6 | ≈ 2.3 | 0.56 |
| $\mu_0 H_{c2}(0)$ [T] | 25 | ≈ 150 | 14 |
| $\xi(0)$ [nm] | 11.5 | ≈ 1.5 | 4.9 |
| $\kappa$ | 3.4 | ≈ 100 | 38 |
| $\lambda(0)$ [nm] | 39 | ≈ 150 | 185 |
| $\mu_0 H_{c1}(0)$ [T] | 0.13 | ≈ 0.03 | 0.018 |

*Table II. Characteristic parameters of the superconductors Nb$_3$Sn, YBa$_2$Cu$_3$O$_7$, and MgB$_2$ (unless otherwise specified, data are for sample A2). Data for Nb$_3$Sn are compiled from Refs. [9] [31] [32] [33] [34] [35]; YBa$_2$Cu$_3$O$_7$ from Refs. [24] [29] [36]; the EDOS for MgB$_2$ is from Ref. [13], tunneling gaps for MgB$_2$ from Refs. [37] [38] [39] [51] ((\*) see also Table III); other data are from this work. $M_{gat}$, mass of one gram-atom (1 mol MgB$_2$ = 3 gat); $V_{gat}$, volume of one gram-atom; $T_c$, average critical temperature obtained by an equal-entropy construction of the idealized specific heat jump; average values of phonon energies $\omega$ as defined in the text; $\gamma$, Sommerfeld constant; N(0), bare band-structure EDOS at the Fermi level; $\lambda_{ep}$, electron-phonon coupling constant; $M\overline{\omega}_2^2$, average lattice force constant; $N(0)<I^2>$, electronic Hopfield parameter; $\Delta C/\gamma T_c$, normalized specific heat jump at $T_c$; $H_c(0)$, thermodynamic critical field at $T = 0$ obtained by*

*integration of the specific heat difference; $E_c$, measured condensation energy (Section 9); $0.2364\gamma T_c^2$, BCS value of the condensation energy in the isotropic, weak-coupling limit; $\Delta(0)$, superconducting gap; $(dH_{c2}/dT)_{T_c}$, slope of the upper critical field at $T_c$, as evaluated from the shift of the onset of the specific heat anomaly in fields from 1 to 3 T; $H_{c2}(0)$, upper critical field at $T = 0$; $\xi(0)$, coherence length at $T = 0$; $\kappa \equiv \lambda/\xi$; $\lambda(0)$, London penetration depth at $T = 0$; $H_{c1}(0)$, lower critical field at $T = 0$.*

The obtained PDOS is a rough estimate, and certainly not unique. However, its low order moments are precise and the Eliashberg equations do not require any more detailed knowledge (assuming uniform coupling $\alpha^2(\omega) \equiv \alpha^2 F(\omega)/F(\omega) = const$). The functional derivative of $T_c$, with respect to variations of the PDOS, has a broad maximum near $2\pi T_c$ [17], whereas the specific heat $C/T^3$ is sensitive to frequencies $\approx 5\,T$. Therefore, in some sense, both $T_c$ and $C/T^3|_{T=T_c}$ exhibit a similar spectral sensitivity to the phonons.

The main result of the "inversion" of the specific heat data, is the determination of average frequencies important for superconductivity, as listed in Table II. The logarithmic average, $\omega_{ln}$ = 480 K, is $\approx$ 4 times larger than the value obtained for the A15 superconductor Nb$_3$Sn, with $T_c$ = 18 K. If the coupling constant $\lambda_{ep}$ of MgB$_2$ was equal to that of Nb$_3$Sn, this high frequency could explain a $T_c$ as high as $\approx$ 70 K. However, the moment $\overline{\omega}_2$ = 633 K is also 4 times larger, so that the average force constant $M\overline{\omega}_2^2$ increases by a factor of two with respect to Nb$_3$Sn. Now assuming an identical Hopfield electronic parameter $\eta$ for both compounds, $\lambda_{ep} = \eta/M\overline{\omega}_2^2$ would decrease to $\approx$ 0.9 in MgB$_2$, bringing $T_c$ back below 30 K. Indeed, the estimate of $\lambda_{ep}$, based on the measured $\gamma$ and a value of $N(0)$ from band-structure calculations [13], is only $\approx$ 0.6, suggesting that in addition $\eta$ has decreased. In other words, as long as we assume that the $\alpha^2(\omega)$ function is smooth and structureless, phonon frequencies alone cannot readily explain the high $T_c$ of MgB$_2$.

## 6. Specific heat near the transition temperature

For a classic BCS superconductor with a Debye lattice, the normal state would appear as a straight line, $C_n/T = \gamma + \beta_3 T^2$, and the electronic term would vanish exponentially at low temperature in the superconducting state, making only the lattice specific heat persist below $\approx T_c/7$. Therefore the initial slopes of both curves, $C_n/T$ and $C_s/T$, should be equal up to a temperature of several Kelvin. Figure 5 shows, for sample A2, the specific heat in the superconducting state for $B = 0$, and the specific heat in the normal state for $B$ = 14 and 16 T (the latter data are practically undistinguishable). The experimental points immediately rise above the BCS result, indicating available low-energy excitations. *This anomaly is the experimental evidence for an unconventional gap.* Note that the EDOS vanishes as $T \to 0$: all electrons are condensed. One notices a weak negative curvature in the normal state specific heat as $T \to 0$, this may have several origins as already mentioned. This negative curvature is also observed, to a lesser extent, in sample HP14, and in Ref. [7]. The Sommerfeld constant is obtained directly from this plot: $\gamma = 0.89 \pm 0.05$ mJ/K$^2$gat (2.7 mJ/K$^2$mol, 1530 erg/K$^2$cm$^3$) for sample A2. The value is slightly less for sample HP14, as could be expected from the $\approx 5\%$ MgO content: $\gamma = 0.83 \pm 0.02$ mJ/K$^2$gat. These values of $\gamma$ lie between those of Cu (0.69 mJ/K$^2$gat) and Al (1.35 mJ/K$^2$gat). Compared to the A15 compounds with highest $T_c$, values having $\gamma \approx 10$ to 25 mJ/K$^2$gat, it is quite small. With $N(0)$ from Ref. [13], the mass enhancement factor is $1+\lambda_{ep} \cong 1.6$, close to the estimate 1.7 of Ref. [13]. There is no reason to expect an exceptionally large Hopfield parameter $\eta = N(0) <I^2>$ in these conditions. Indeed, if $\eta$ is estimated as $\lambda_{ep} M \overline{\omega}_2^2$, where $\lambda_{ep}$ is taken from the mass renormalization, one finds two third of the value of $\eta$ for Nb$_3$Sn (Table II). This implies that, in the framework of Eliashberg theory with isotropic coupling and isotropic gap, neither phonon frequencies, nor the EDOS can account for the high $T_c$ of MgB$_2$.

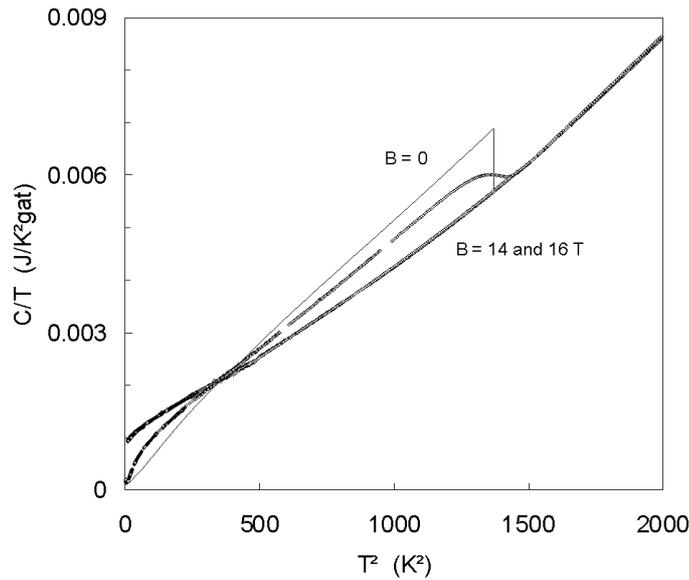

*Figure 5.* Normal-state specific heat of sample A2 for B = 14 and 16 T, and superconducting-state specific heat for B = 0. The thin line is a BCS curve with $T_c$ = 37 K and $\gamma$ = 0.89 mJ/K$^2$gat.

The specific heat jump at $T_c$ appears rounded in Fig. 5, this is partly because of the large slope of the lattice background. After removing this background (obtained by averaging the data measured in 10 and 14 T), the specific heat jump is better defined (Fig. 6). Samples A2 and HP14 do not differ much in this respect. We do not doubt that the sharpness of the transition will improve in future samples (see e.g. Ref. [6]); however, on a relative scale, this result is sufficient for thermodynamic studies. The data in fields of 0, 0.5, 1, 2 and 3 T were measured with the adiabatic calorimeter, at the low end of the useful range of Pt thermometry (the thermometer used in this calorimeter was calibrated in fields up to 16 T and down to 15 K, using a dielectric thermometer to maintain a constant temperature during field changes).

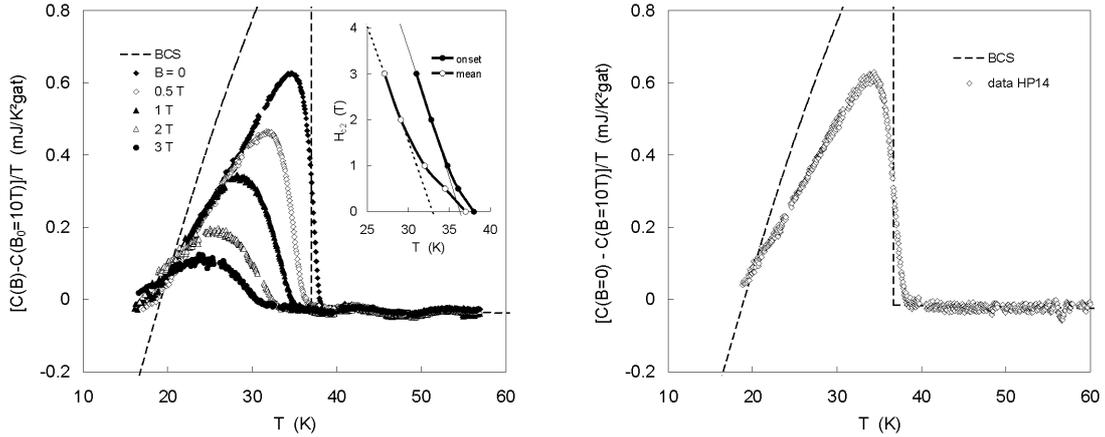

*Figure 6. (a), left panel: sample A2, specific heat difference between the superconducting and normal state for B = 0, 0.5, 1, 2, and 3 T (adiabatic calorimetry). The normal state is a smoothed average of the B = 10 and 14 T curves. **Inset**: phase diagram in the H-T plane showing the line of transition onsets (●) and the midpoints (○). (b), right panel, same plot for sample HP14.*

The normal- and superconducting- state curves intersect at 19 K, close to $0.5 T_c$. This crossing is predicted to occur at $0.52 T_c$ in the isotropic weak-coupling BCS limit [40]. Superconductors with $\lambda_{ep} \approx 1$ tend to follow the two-fluid model, $(C_s - C_n)/\gamma T_c t = 3t^2 - 1$, $t \equiv T/T_c$, with a crossing near $0.6 T_c$ [41].

The transition is shifted to a lower temperature in bulk by the application of a magnetic field. In a high-$T_c$ superconductor, the transition onset would remain constant, the transition width would increase, and the height of the peak would vanish rapidly as the field is increased [29]. In a classic isotropic type-II superconductor (e.g. $Nb_{77}Zr_{23}$, [41]), the shape of the transition would remain essentially unchanged, and the bulk transition would be shifted to lower temperatures. In the present case, the *onset* of the transition does remain well defined, but the width of the transition increases abnormally. The same observation was made in resistive determinations of the upper critical field, which unlike specific heat may be broadened by flux-flow effects [25] [42]. If the superconductor is not in the clean limit, local variations of the mean free path might broaden the $H_{c2}(T)$ line at high fields. A more likely explanation is the anisotropy of the upper critical field. Grains are randomly oriented in the present polycrystalline sample. If the critical field is a function of the angle θ between $\vec{H}$ and the direction normal to the boron planes, $T_c$ of the grains is distributed between the values defined by the $H_{c2}(T, H // ab)$ and $H_{c2}(T, H \perp ab)$ lines.

4The width of the distribution evidently increases with *H*. In this model, the onset of the specific heat jump corresponds to the $H_{c2}(T)$ line in the "hard" direction, presumably *H* parallel to the boron planes. This picture is confirmed by recent measurements on single crystals, which give an anisotropy $H_{c2}(H//ab)/H_{c2}(H\perp ab) = 2.6$ to 2.7 [43] [44]. We have plotted this onset in the *H* - *T* phase diagram (Fig. 6a, inset). The slope of this critical field is –0.56 T/K. Assuming a WHH [45] temperature dependence, we extrapolate this initial slope to a zero-temperature value close to 14 T. This value is consistent with our finding that the specific heat at low temperature is the same for both 14 and 16 T. In the same phase diagram, we also show the shift of the midpoint of the calorimetric transition. These values of $T_{c2}(H)$ extrapolate to 11.5 T at *T* = 0 (WHH), and probably represent an angular average of $H_{c2}(T,\theta)$. However, the latter extrapolation is affected by an uncertainty due to the positive curvature at low fields. Such a curvature has been observed in some classical superconductors, both isotropic (e.g. $Nb_3Sn$ [34]) and anisotropic (e.g. $NbSe_2$ [46]). Fermi surface and pairing anisotropies could produce these effects [47] [48] [49] [50].

Transport measurements of $H_{c2}(T)$ were performed, using hot-pressed samples prepared with the same initial powders, up to *B* = 17 T [25]. The midpoints of the resistive transitions follow the midpoints of the specific heat jumps at low fields, with the same positive curvature, but the high field $H_{c2}(T)$ curve is more linear than predicted by WHH, giving $H_{c2}(0)$ = 14 T as obtained above. Similar results are obtained by Ref. [42]. One should keep in mind that these results are angle-averaged, as recent single crystal measurements up to 9 T have revealed an $H_{c2}$ anisotropy factor of 2.6 [43].

Independently of its origin, the fast decrease of the specific heat jump is required by thermodynamics. We show below that the specific heat $C/T|_{T\to 0}$ increases rapidly as a function of *H*, at a rate that is much higher than the usual $\gamma_m(H) \cong \gamma H/H_{c2}$. Therefore, in order to preserve the entropy balance, the amplitude of the specific heat jump has to decrease at an unusually high rate. In this sense, the suppression of the jump is related to the small critical field associated with one of the gap components (see Section 7).

The dashed line in the main panel of Fig. 6a shows the BCS line based on the measured $T_c$ and γ [40]. It grossly overestimates the jump, as also seen in Fig. 5. The dimensionless jump ratio, $\Delta C/\gamma T_c$, is only found to be 0.82 ± 10%. Gap anisotropy is known to lower the specific heat jump, essentially because electrons on the Fermi surface condense at a lower rate. Alternatively, strong coupling would cause a deviation in the *opposite* direction, and some compensation may be present.

## 7. Specific heat at low temperature: anomalous gap

The specific heat of sample A2 was measured at low temperatures (2-20 K in zero field, 3-16 K in *B* > 0) in a second calorimeter, using a modified relaxation technique [5]. The extra contribution of the addenda never exceeded 25% of the total heat capacity. Fields $B$ = 0, 1, 3, 14 and 16 T were applied in the normal state before cooling through the transition in order to avoid metastable states. As already mentioned, the 14 and 16 T data coincide within experimental accuracy, giving $\mu_0 H_{c2}(0) \lesssim 14$ T. These normal-state data, already included in Fig. 5, determine that $\gamma = 0.89 \pm 0.05$ mJ/K$^2$gat. This result agrees with Refs. [3] [6] and [7], who used magnetic fields up to 9 T to suppress most of the superconductivity; it differs from Ref [4], who used an extrapolation of zero-field measurements.

The specific heat difference $(C_s - C_n)/T$ versus $T$ is plotted in Fig. 7a for sample A2. The normal-state data $C_n(T)$ were smoothed before subtraction. The data in 14 and 16 T from the relaxation calorimeter up to 22 K, and the data in 10 and 14 T from the adiabatic calorimeter down to 20 K are also included in this plot after subtracting their smoothed values, in order to show the scatter of the points. The data for sample HP14 (Fig. 7b) were obtained by point-to-point subtraction. The BCS curve (full line), based on the measured value of $\gamma$, is clearly inconsistent with the low temperature data. The upturn at low temperature in the $B$ = 0 curve is due to the Curie-Weiss paramagnetism of impurities.

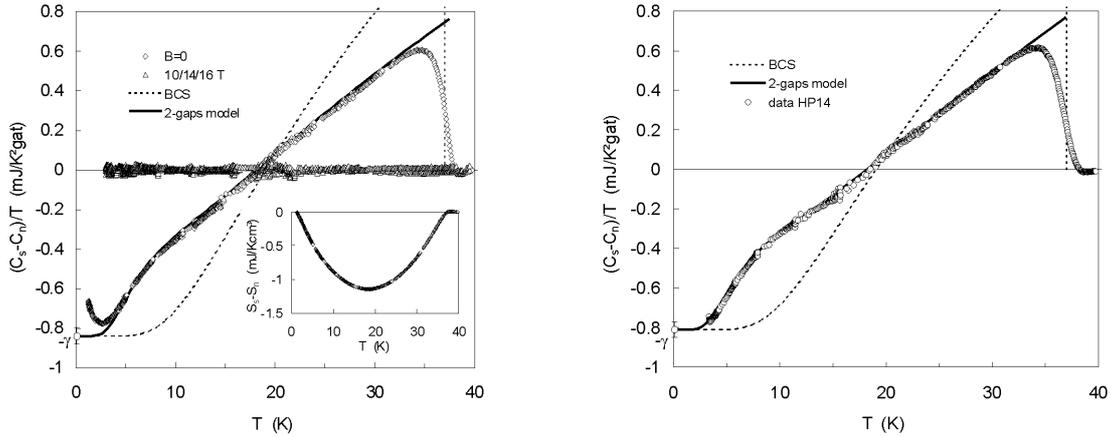

*Figure 7. (a), left panel*: (◇), *specific heat difference between the superconducting state (B = 0) and the normal state versus temperature for sample A2. The normal state is a smoothed curve based on the average of the data in B = 10 T (15 < T < 40 K), 14 T (3 < T < 40 K) and 16 T (3 < T < 16 K).* (△), *residual differences between the high field data and the smoothed normal-state curve. Both adiabatic (T > 16 K) and relaxation (T < 22 K) calorimetry are included. Dotted line: BCS curve. Full line through the data: two-gap model with the parameters of Table III, $\gamma = 0.84$ mJ/K$^2$gat and $T_c$ = 38 K.* ***Inset**: entropy difference obtained by integration of $(C_s - C_n)/T$ from 39 K downwards.* ***(b), right panel***: (○), *specific heat difference for sample HP14. Dotted line: BCS curve. Full line: two-gap fit with the parameters of Table III, $\gamma = 0.81$ mJ/K$^2$gat and $T_c$ = 37.5 K.*

The net area under the $(C_s - C_n)/T$ versus $T$ curve is the entropy difference between the normal and the superconducting state. By integration of the data from 39 K to a variable temperature $T < T_c$, we obtain the curve shown by symbols in the inset of Fig. 7a. We have performed no correction for the paramagnetic contribution of impurities, neither for $B = 0$, nor for $B = 14$ or 16 T; the magnetic entropy is present in both fields and partly cancels out. The fact that the entropy difference heads toward zero as $T \to 0$ is important as it asserts the thermodynamic consistency of the data, *including the deviation from the BCS curve at low temperature.*

This deviation is one of the central points of the present study. Qualitatively, one may consider that the specific heat difference shown in Fig. 7 results from the superposition of two components. One is a BCS superconductor representing approximately half the volume of the sample. The other component appears to remain in the normal state above $T_c/2$. Then at lower temperature, a superconducting gap seems to open progressively for this second component, so that the specific heat difference $(C_s - C_n)/T$ decreases until it reaches $-\gamma$ at $T = 0$ (after correction for the magnetic contribution for sample A2). Note that there is not two specific heat jumps. This second gap does not close at a lower critical temperature, but at $T_c$.

In order to reproduce the peculiar behavior of the superconducting state specific heat, we tested several models (A-C). They all have to go beyond the image of a single, isotropic BCS gap. This finds support in the wide spread of values reported for $\Delta(0)$ in tunneling experiments [37] [38] [39] [51].

(A) Shulga *et al.* pointed out that the rather high value of the upper critical field $H_{c2}(0) \cong 14$-18 T sets constraints on band and coupling scenarios [14]. They proposed to divide the carriers on the Fermi surface into two subgroups, one of them with heavy masses and strong coupling to low frequency phonons, the other one with weak coupling to high frequency phonons. The $\alpha^2 F(\omega)$ function has two peaks near 60 and 17.5 meV. This finally results into two gaps with $2\Delta_1(0)/k_B T_c = 5$ and $2\Delta_2(0)/k_B T_c = 3$, respectively. We are indebted to E. Schachinger who used the relevant parameters of this model to calculate the thermodynamics by numerically solving the Eliashberg equations [52]. Although some integral features, such as the specific heat jump and the condensation energy, could be accounted for, the model failed to reproduce the peculiar low-energy excitations. For this, much lower gap values are needed. It is possible to obtain a better agreement by moving the low frequency phonon to about 8 meV, with a corresponding decrease of $\Delta_2(0)$ [53]. Up to now, no phonon peak has been detected in this frequency range, neither in the lattice specific heat (see Section 5) nor in neutron experiments [30].

(B) We have calculated the specific heat in the low temperature limit $\Delta(T) \cong \Delta(0)$ for an extended *s*-wave gap with uniaxial anisotropy, $\Delta(\vec{k}) = \Delta_{\max}(1 - e\cos^2\theta)$. With $e = 1$, this model yields $C(T \to 0) \propto T^2$, accidentally similar to the *d*-wave case (Fig. III.18 of Ref. [29]). On the average, this reproduces crudely one of the features of the measurements, i.e. an excess specific heat with respect to the isotropic BCS curve, but it overestimates the specific heat below 5-10 K where the rapidly vanishing specific heat suggests a finite gap. A finite anisotropy ($e < 1$) would probably be more appropriate [54] [55], but such continuous distributions of $\Delta(\vec{k})$ generally tend to produce specific heat curves which have much less structure than the measured ones.

(C) A two-gap structure where both gaps close at $T_c$ is suggested by Raman- [56], photoemission- [57], point-contact- [58] and tunneling- spectroscopy [59], and was proposed theoretically by two groups [14] [15]. The specific heat can be calculated easily under simplifying assumptions: the variation of the gaps with the temperature is identical to that of a single BCS gap, i.e. $\Delta_i(T)/\Delta_{BCS}(T) = \Delta_i(0)/\Delta_{BCS}(0)$, and the contributions are linearly additive with suitable weights $\gamma_i$. Technical details are given in Appendix I. The fits are shown in Fig. 7 for samples A2 and HP14, and the parameters are summarized in Table III. A similar fit to another sample [6] [60] is presented in another article of this volume [8], and the values are incorporated in Table III along with estimations recently obtained by spectroscopic techniques. All agree on $\Delta_1(0)/\Delta_2(0) \approx 3$. In the classification of Liu *et al.* [15], $\Delta_1(0)$ originates from 2D hole sheets, and $\Delta_2(0)$ from 3D $p_z$ sheets on the Fermi surface. The essential point made by the present specific heat experiments and analyses is that this two-gap scenario finds support in *bulk* thermodynamics. The larger gap has a typical BCS-like value, showing moderately strong coupling. The lower gap represents a "weak" superconducting channel. Alone, it would justify superconductivity at no more than $\approx 10$ K. We proceed to demonstrate that this second channel gives rise to a "partial" critical field that is small compared to $\mu_0 H_{c2} \cong 14\text{-}16$ T.

*Table III. Gap parameters from specific heat and spectroscopic experiments. (*), see Section 10.*

|  | Technique | $2\Delta_1(0)/k_B T_c$ | $2\Delta_2(0)/k_B T_c$ | $\gamma_1 : \gamma_2$ |
|---|---|---|---|---|
| This work, sample A2 | specific heat | 3.8 | 1.3 | 0.5:0.5 |
| This work, sample HP14 | specific heat | 3.9 | 1.3 | 0.5:0.5 |
| Ref. [6], [8], [60] | specific heat | 4.4 | 1.2 | 0.55:0.45 |
| Ref. [61] (*) | penetration depth | 4.6 | 1.6 | 0.60:0.40 |
| Ref. [59] | tunneling | 4.5 | 1.9 |  |
| Ref. [56] | Raman | 3.7 | 1.6 |  |
| Ref. [58] | point-contact | 4.1 | 1.7 |  |
| Ref. [57] | photoemission | 3.6 | 1.1 |  |
| Ref. [15] | band structure | 4 | 1.3 | 0.53:0.47 |

## 8. Field dependence at low temperature

For a classic isotropic s-wave superconductor (e.g. $Nb_{77}Zr_{23}$ [41]), the specific heat at $T \ll T_c$ is dominated by the vortex cores [62] [63], the number of which is proportional to the field. As a result, the linear term of the electronic specific heat in the mixed state increases linearly with the field, $\lim_{T \to 0} C_s/T \equiv \gamma_m(H) = \gamma H/H_{C2}(0)$ [64].

The low temperature specific heat of sample HP14 for $B$ = 0, 0.05, 0.1, 0.2, 0.3, 0.5, 1, 2, 4, 8, and 14 T is shown in Fig. 8. Measurements for sample A2 for $B$ = 0, 1, 3, 14 and 16 T are shown in the inset. In the presentation $C/T$ versus $T^2$, the lattice contribution essentially adds a positive slope to the intercept $\gamma_m(H)$ at $T = 0$. These plots show two anomalies. First the $C/T$ data in $B = 0$ are not linear in $T^2$. This is the anomalous contribution of $\Delta_2(T)$ discussed in Section 7. The second anomaly is the *non-linear* increase of the specific heat with the field at low temperature, as shown by the plot of $[C(B,T) - C(0,T)]/T$ at $T = 3\,\text{K}$ on a logarithmic field scale (Fig. 9). The saturation of $C/T$ above $\approx H_{c2}(0)/2$, and the crossover to a different low-field behavior near 10 K, are also intriguing.

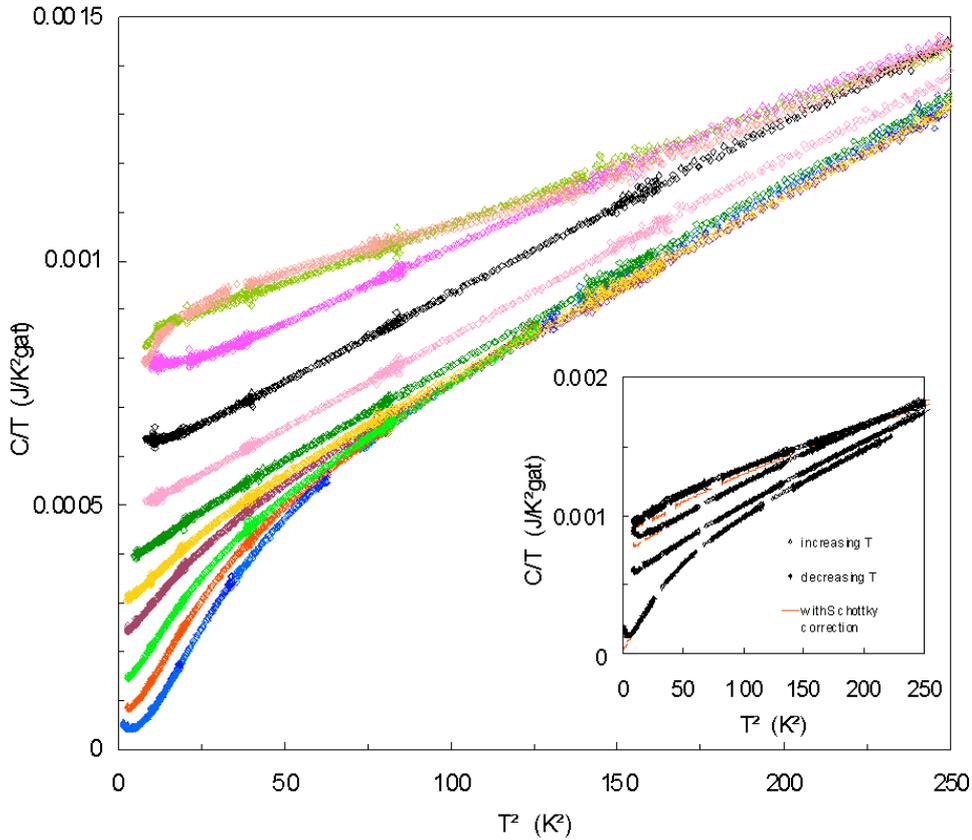

*Figure 8.* Main panel: total specific heat of sample HP14 below 16 K in a C/T versus $T^2$ plot. From bottom to top: B = 0 (blue), 0.05, 0.1, 0.2, 0.3, 0.5 (green), 1, 2, 4, 8 (lemon green), and 14 T (coral pink). Note the non-linearity of the curve for B = 0 and the non-linear field dependence for a given temperature. **Inset**: same plot for sample A2 for B = 0, 1, 3 14 and 16 T. The data for 14 and 16 T coincide within experimental error. Full symbols are measured with decreasing temperature, open symbols with increasing temperature. The red lines are corrected for the magnetic Schottky contribution of 120 ppm (atomic) S = ½ spins (Appendix II).

A theory for such an extraordinary field dependence in a fully gapped superconductor is not available. The qualitative image that emerges is that of a smaller "weak" gap $\Delta_2$, that is washed out by a field of the order of 0.5 T; at this point the $C/T$ curve becomes almost linear in $T$ (Fig. 8), and the sample appears as a superposition of a normal metal with $\gamma_2 \approx \gamma/2$ coexisting with a high field superconductor. A critical field of the order of 0.5 T for the weak component is not uncommon for a 3D, type-II superconductor having similar gap values, e.g. $\mu_0 H_{c2} \approx 0.4$ T for pure Nb.

Although this picture is appealing, it does not explain the quasi-logarithmic field dependence of the density of states as $T \to 0$, which persists above 0.5 T. Two gaps and two Fermi velocities define two coherence lengths, and one can ask what is the shape and size of a vortex core in these conditions. The topology of vortex cores above 0.5 T could give rise to new phenomena, just like the existence of nodes dramatically affects the low-temperature specific heat of *d*-wave superconductors [65] [66].

The contribution of paramagnetic Fe impurities cannot trivially explain these peculiar features. For sample A2, a correction for the Schottky contribution (see Appendix II) was applied and is shown in the inset of Fig. 8 by red lines just below the data; the main effect occurs for the zero-field curve below 3 K. The magnetization curves of sample HP14 have evidenced a smaller concentration of magnetic moments. Accordingly, the specific heat of sample HP14 can be measured down to lower temperatures without showing excessive upturns, and any reasonable correction for the residual Schottky contribution leaves the characteristic gap features unaffected.

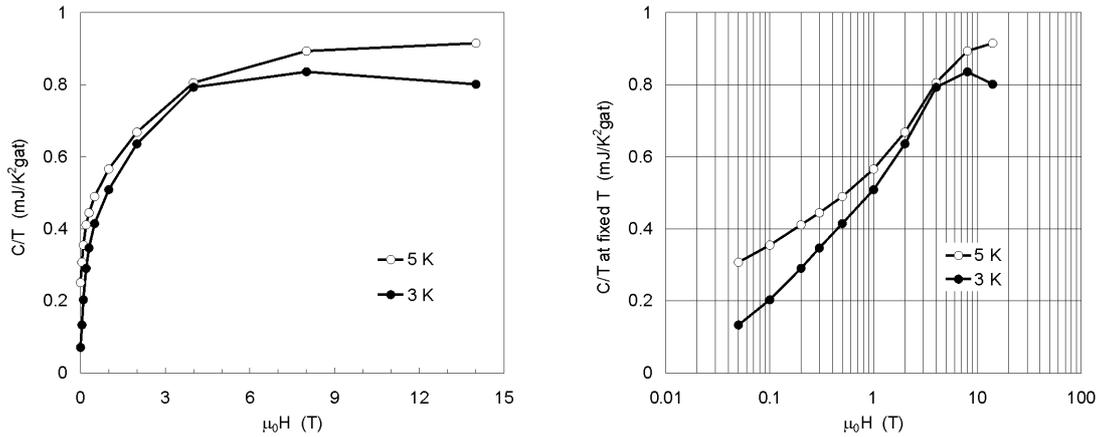

*Figure 9.* Specific heat C/T of sample HP14 at 3 and 5 K versus the magnetic field on a linear scale (left panel) and a logarithmic scale (right panel). Note that $B_{c1} < 0.05$ T. The maximum at 8 T at 3 K is believed to be an artifact resulting from the residual Schottky contribution for 4 and 8 T and the proximity of $T_{c2}(H)$ at 14 T.

## 9. Ginzburg-Landau parameters

We calculate the thermodynamic critical $H_c(T)$ field according to (cgs units):

$$-\frac{H_c^2(T)}{8\pi} \equiv \Delta F(T) = \Delta U(T) - T\Delta S(T)$$

$$\Delta U(T) = \int_T^{T_c} [C_s(T') - C_n(T')] dT'$$

$$\Delta S(T) = \int_T^{T_c} \frac{C_s(T') - C_n(T')}{T'} dT'$$

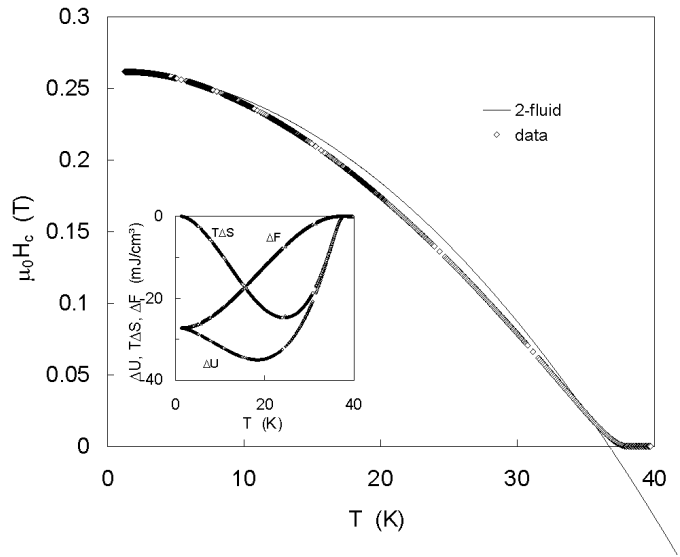

**Figure 10.** ($\diamond$): *thermodynamic critical field curve obtained by integration of $C_s - C_n$. Full line: reference two-fluid dependence $h = 1 - t^2$ ($T_c$ is defined by an equal entropy construction). **Inset:** internal energy difference $\Delta U$, free energy difference $\Delta F$ and entropy difference multiplied by the temperature $T\Delta S$, versus temperature. The condensation energy tends to 27 mJ/cm$^3$ at $T = 0$ (sample A2).*

The result for $H_c(T)$ is given in Fig. 10 and Table II. The inset of Fig. 10 shows separately the differences of internal energy ($\Delta U$), entropy multiplied by the temperature ($T\Delta S$), and free energy ($\Delta F$). The thermodynamic field reaches 0.26 T at $T = 0$. The corresponding condensation energy is one-quarter of the value for Nb$_3$Sn, and one-sixteenth of the value for YBa$_2$Cu$_3$O$_7$ (Table II). MgB$_2$ stands out as a high temperature superconductor with a very low EDOS. Part of the condensation energy is saved by anisotropic pairing.

The type-II characteristic parameters listed in Table II are obtained from the Ginzburg-Landau relations:

$$H_{c2} = \sqrt{2}\kappa H_c = \frac{\Phi_0}{2\pi\xi^2}$$

$$H_{c1} = \frac{H_c}{\sqrt{2}\kappa}\ln\kappa = \frac{\Phi_0}{4\pi\lambda^2}\ln\kappa$$

where $\Phi_0 = 2.07\times10^{-7}$ Gcm$^2$ is the flux quantum. The second equation is valid for large $\kappa$; here $\kappa = 38$. At $T = 0$, the coherence length is 49 Å and the London penetration depth 1800 Å. The corresponding lower critical field, 180 G, is not apparent in the magnetization curves, and may be underestimated; the ZFC diamagnetism starts to deviate from $\chi = -1/4\pi$ at the 1% level between 300 and 500 G (Fig. 2a; the full magnetization curves are found in Ref. [5]). An *a posteriori* verification shows that all specific heat curves in B > 0 were measured in the mixed state.

## 10. Conclusion

Normal- and superconducting-state properties of MgB$_2$ were characterized by thermodynamic measurements on two samples, with similar results. Many characteristic features differ strongly from those of other superconductors:

- $\lambda_{ep}$ estimated from the *isotropic* Allen-Dynes equation on the one hand, and $\lambda_{ep}$ estimated from the renormalization of the EDOS on the other, differ by a factor of nearly two;

- the normalized curve of the thermodynamic critical field is much closer to the BCS limit than to the strong-coupling curve that is generally found for classic "high" $T_c$ superconductors. In terms of the deviation function, $D(t) \equiv H_c(T)/H_c(0) - (1-t^2)$ is *negative* and BCS-like, whereas it is usually zero or positive for classic superconductors with $T_c$ in the 10-20 K range owing to strong coupling corrections;

- the normalized specific heat jump $\Delta C/\gamma T_c \cong 0.8$ is unusually small;

- the condensation energy is low;

- in a magnetic field, the onset of the calorimetric transition remains well-defined whereas the bulk transition is broadened, suggesting an angular dependence of $H_{c2}$;

- the specific heat at $T \ll T_c$ and zero field shows a relatively large extra weight with respect to that of a BCS superconductor;

- the specific heat at $T \ll T_c$ in a field is also essentially different from that of an isotropic s-wave superconductor, with an overall logarithmic rather than linear field dependence. One observes two distinct regimes in the field dependence for $T < 10$ K and $T > 10$ K.

These anomalies in zero field are well accounted for by the two-gap model proposed here. The larger gap is BCS-like, and the smaller gap is about three times smaller. Both

23close at $T_c$. They contribute to the thermodynamics with approximately the same weight. These features confirm with success the predictions of Liu *et al.* based on first-principle calculations of the band structure, phonon frequencies and electron-phonon coupling [15]. The main result of this study is the consistency found between two-gap models obtained by spectroscopic methods, which are surface sensitive characterizations, and the two-gap model that consistently fits *bulk* specific heat data for several samples. Other bulk probes can be used to further test this model. At low temperature, the penetration depth, $\lambda^{-2}(T)$, is a measure of the condensate density, whereas the electronic specific heat $C/T$ is a measure of the density of excitations. They are complementary. Therefore the $\lambda^{-2}(T)$ curve should mirror the $C/T$ curve of Fig. 7 or Fig. 11. This is indeed observed in recent data of Manzano *et al.* [61], which we can fit with similar gaps values and weights as used here (Table III).

This simple model neglects any coupling between the gaps (except that coupling ensures that both gaps close at the same temperature), and treats the two-band superconductor as two independent systems. The reason why it accounts so well for thermodynamic properties is not entirely clear. We hope that these experimental results will stimulate further theoretical investigations, in particular in non-zero fields where unusual and significant effects remain unexplained. Alternative scenarios, such as a change in the symmetry of the order parameter at some temperature in the vicinity of 10 K, are under investigation [52]. In any case, the present results obtained from polycrystalline samples strongly motivate the study of single crystals.

**Acknowledgements**

We thank B. Revaz for the development of the relaxation calorimeter, M. Decroux, J.-Y. Genoud and E. Walker for test samples and communication of their unpublished data, E. Schachinger and S.-L. Drechsler for scientific discussions and preliminary calculations, and T. Tybell for help in the finalization of the manuscript. This work was supported by the Fonds National Suisse de la Recherche Scientifique.

**Appendix I: specific heat in a two-gap model**

We refer to a forthcoming manuscript that will be published jointly by the Berkeley and Geneva groups [60]. The two-gap superconductor is treated as a superposition of two independent subsystems with gap ratios $2\Delta_1(0)/k_BT_c$ and $2\Delta_2(0)/k_BT_c$. A calculation of the specific heat in the superconducting state for variable $2\Delta(0)/k_BT_c$ is given in Ref. [23]. Although the original treatment given by Padamsee *et al.* refers to strong-coupling with $2\Delta(0)/k_BT_c \geq 3.5$, the same formalism can be used for arbitrary ratios as long as the gap function has the same variation with temperature as in the BCS theory. The analysis proceeds from the formula for the entropy of a system of independent fermions $S = -k_B \sum [f \ln f + (1-f)\ln(1-f)]$, where $f$ is the Fermi function, and where the quasiparticle spectrum is taken to be the same as in the BCS (weak-coupling) theory $E_k^2 = (\varepsilon_k^2 + \Delta^2)$.

Following Ref. [23], we generated a set of normalized functions $C/\gamma T_c t = \partial(S/\gamma T_c)/\partial t$ versus $t \equiv T/T_c$ for $1 \leq 2\Delta(0)/k_B T_c \leq 5$ (see e.g. Fig. 11). The curve for $2\Delta(0)/k_B T_c = 3.5$ has the well-known BCS shape [40]. The curves for very low values of $2\Delta(0)/k_B T_c \approx 1$, never investigated before, show a rapid increase of the specific heat when the thermal energy $k_B T$ becomes of the order of the gap $\Delta(T) \ll k_B T_c$, a regime that does not occur in the BCS theory. The specific heat is then approximately that of a semiconductor with a constant gap. Above this crossover, $k_B T > \Delta(T)$, the specific heat approaches the normal state specific heat. At $T_c$, there is a small jump caused by the closing of the gap; it is small because the superconducting ground state is already essentially empty. The amplitude of the residual jump is proportional to $[\Delta(0)/k_B T_c]^2$ [67]. The entropy balance is preserved. For larger values of $2\Delta(0)/k_B T_c$, and in particular in the BCS limit, thermal excitation of the carriers across the gap occurs mainly in the vicinity of $T_c$.

In order to fit the two-gap model we add two such curves with suitable weights. The hump in the measured specific heat near 10 K is due to the thermal excitation of carriers across the smaller gap; it is necessarily a smooth feature. The jump at $T_c$ is due in majority to the closing of the larger gap $\Delta_1$. The smallness of the overall specific heat jump ratio $\Delta C/\gamma T_c \cong 0.8$ is in this model due to the reduced weight ($\approx 50\%$) attributed to the larger gap; it is nevertheless a nearly BCS-like gap with a moderately strong coupling correction.

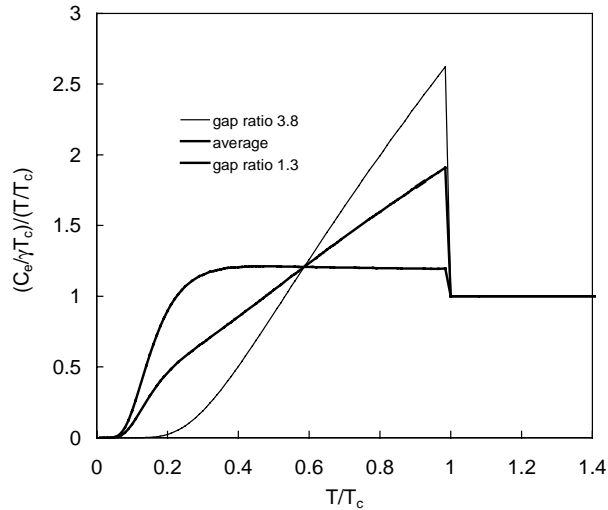

*Figure 11.* Normalized specific heat curves $C/\gamma T_c t$ versus $t \equiv T/T_c$ for two gap ratios $2\Delta(0)/k_B T_c = 1.3$ and $2\Delta(0)/k_B T_c = 3.8$. These two curves, added with equal weights ("average"), fit the specific heat of sample A2 (Fig. 7a).

**Appendix II: corrections for Schottky contributions**

The amplitude of the Schottky anomaly for sample A2 was determined at $B = 3$ T by using $C_{magn}(B,T) = nRx^2 e^x (e^x - 1)^{-2}$, where $x \equiv 2\mu_e B / k_B T$, $\mu_e$ is the electron magnetic moment, and $n$ the atomic concentration of these moments, if $C$ is expressed per gram-atom. $S = 1/2$ is assumed for simplicity. We find that with $n = 120$ ppm, the corrected 3 T curve recovers the expected shape at $T \to 0$. Larger values would cause a sharp unphysical downturn, and would be definitely inconsistent with the Curie susceptibility measured above $T_c$. Keeping this concentration, we can determine the *effective* interaction field $B_{int}$ that causes the upturn when the external field is zero. The zero field curve, measured down to 2 K and corrected for this magnetic contribution, extrapolates to zero if $B_{int}$ is of the order of 1 T. This procedure does not have to be iterated since the effective field $[(3 \text{ T})^2 + (1 \text{ T})^2]^{1/2}$ does not differ much from the applied field at 3 T. The 1 T curve, which is measured only above 3 K, is virtually unaffected by this correction. The 14 and 16 T curves are affected, and lose some of their negative curvature; however, the raw data and the corrected curve point almost to the same value of γ at $T = 0$ (Fig. 8, inset). For simplicity, we did not subtract this magnetic contribution before inverting the specific heat to find the phonon frequencies (Fig. 4). The PDOS at the lowest energy that came out of the inversion is possibly in part an artifact due to this contribution, since the normal state curves below $T_c$ are in fact the high field curves. However, the corresponding Einstein mode has such a low weight, 0.01% of the total number of modes (Table I), that it does not affect the determination of the moments of the PDOS. A correction was applied only in the inset of Fig. 8, red curves. For sample HP14, such impurity contributions are less important, and therefore no corrections were applied.